\documentclass[12pt,preprint]{revtex4}
 \usepackage[T1]{fontenc}
 \usepackage[utf8x]{inputenc}
\usepackage{amsmath}
\usepackage{amsfonts,amssymb}
\newcommand{\sign}{\text{sign}}

\def\be{\begin{equation}}
\def\ee{\end{equation}}
\def\ba{\begin{eqnarray}}
\def\ea{\end{eqnarray}}

\begin{document}
\bibliographystyle{plainnat}

\title{Exact solutions in two-dimensional metric $f(R)$-gravity}

\author{Maria Shubina }

\email{yurova-m@rambler.ru}

\affiliation{Skobeltsyn Institute of Nuclear Physics\\Lomonosov Moscow State University
\\ Leninskie gory, GSP-1, Moscow 119991, Russian Federation}


\begin{abstract}

In this paper we consider the two-dimensional metric $f(R)$-gravity model for the metric tensor depending on two variable: time and one spacelike coordinate. We obtain exact analytical vacuum solutions for different forms of function $ f(R) $ which are solutions of cosmological type. These solutions are expressed in terms of arbitrary functions, which, under certain conditions, can be chosen as new variables.

\end{abstract}

\keywords{$f(R)$-gravity, exact solution }

\maketitle

\section{Introduction}

The $f(R)$-gravity is one of the popular modifications of Einstein's general relativity, in which the action is described by a function $ f(R) $ of the Ricci scalar $ R $, see \cite{SF}-\cite{NO} and references therein. Interest in modified theories of gravity continues unabated today. The motivation for this associated with cosmology, astrophysics, high energy physics remains relevant. Due to $f(R)$-gravity model consists of a system of coupled nonlinear partial differential equations, the finding of exact solutions is a very difficult task. However, much progress has been made in constructing of exact solutions depending on one variable. Also in the work \cite{GPLM} static axially symmetric vacuum solutions were found including for $ f(R) = k  \sqrt{R}$. We will discuss this in more detail below. For $ f_{R} = R^{1 + \delta} $ a class of exact static spherically symmetric solution has been obtained in \cite{CB}. In \cite{MV} authors construct the spherically symmetric solutions of using their input function method. In \cite{SZ} a Lagrangian derivation of the equations of motion for the vacuum static spherically symmetric metrics in $ f(R) $-gravity is presented. It should be noted that in the last two papers a solution with $ f(R) \sim \sqrt{R}$  is also present. Also $ \sqrt{- R} $ is in \cite{CFV} where some new static vacuum spherically symmetric solutions are found with the method of Noether symmetries. The static cylindrically symmetric vacuum solutions is constructed in \cite{AMN}. In \cite{GS} the authors find a new method for obtaining static and spherically symmetric solutions in $ f(R) $ theory. Note that among the functional dependences of $ f(R) $ on $ R $ obtained in this work there are $ f(R) \sim \sqrt{R} $ and $ f(R) \sim \frac{1}{R} $. There are also a number of works in which exact solutions of the equations of gravity with matter fields are constructed, mainly depending on one variable, see for example \cite{HL} -- \cite{VIP}.

In this paper we consider the two-dimensional metric $f(R)$-gravity model in which all functions depend on time and one spacelike coordinate. We obtain exact analytical vacuum solutions for different forms of function $ f(R) $ which are solutions of cosmological type. Since we solve all the equations and find solutions in terms of the light-cone coordinates, the results can be extended to the case of axial symmetry. Therefore, it is not surprising that in our consideration, as well as in \cite{GPLM}, the dependence of function $ {f}^{2}(R) \sim R $ as a solution of equations appeared. Further, like \cite{GPLM}, we needed simplifying assumptions to find the exact solutions, namely, in \cite{GPLM} function $ \frac{df(R)}{dR} (\rho, z) = U(\rho) V (z)$ while we have $ \frac{df(R)}{dR} (\zeta, \eta) = {f_{R}}_{1}(\zeta){f_{R}}_{2}(\eta) $. Since they are not the same thing, we can expect that our solutions will differ from those obtained in \cite{GPLM} and it will be interesting to check this in the future. 

This article is organized as follows. In Section 2 we introduce the notation and present the equations of the model under consideration, which we will solve below. In Section 3 we are considering two types of metric interval $ ds^{2} = g_{\mu\nu} dx^{\mu} dx^{\nu} $. For $ g_{\mu\nu} = diag(F, -F, -1, -\alpha^{2}) $ we obtain the expression $ f(R)\sim \sqrt{|R|} $, and for $ g_{\mu\nu} = diag(F, -F, -|\alpha|, -|\alpha|) $ the function $ f(R)\sim |R|^{\kappa} $, $ \kappa \in [\frac{1-\sqrt{3}}{2}, \frac{1+\sqrt{3}}{2}]$ and $ f(R) \sim \frac{1}{R} $. It is shown that it is possible to choose new coordinates so that $ F $ becomes equal to $ 1 $. For $ f(R) \sim \frac{1}{R} $ we obtain a class of solutions parameterized by two arbitrary functions; with an appropriate choice of new variables the metric can be made conformally flat. We also indicate how the off-diagonal components of the spatial part of the metric ($ g_{xy} $) can be obtained.

\section{Models under consideration and field equations}

In this paper we consider the metric $f(R)$-gravity model for the case when the metric tensor depends on two variable, on time and one spacelike coordinate. The gravitational action without matter fields is 
\be
S = \frac{1}{16 \pi G} \int d^{4}x \sqrt{-\textit{g}} \, f(R),
\ee
where $ G $ is the gravitational constant, $ \textit{g} $ is the determinant of the metric tensor $ g_{\mu\nu} $, $ R = g^{\mu\nu} R_{\mu\nu} $ is the scalar curvature, or the Ricci scalar, $ R_{\mu\nu} $ is the Ricci tensor.
Variation of eq.(1) with respect to the metric gives the field equations \cite{SF}
\be
f_{R}(R) R_{\mu \nu} - \frac{1}{2}f(R)g_{\mu \nu} - [ \nabla_{\mu} \nabla_{\nu} - g_{\mu \nu} \square ] f_{R}(R) =  0,
\ee
where $ f_{R}(R)\equiv \dfrac{df(R)}{dR} $.

Further, we write down the metric and equations in the notation similar to those introduced in the work of Belinski\v{i} and Zakharov \cite{BZ}. Thus, the difference between the vacuum equations of general relativity and the equations of the considered model of $ f (R) $-gravity becomes immediately obvious. 

We will consider a metric that depends on time and one spacelike coordinate, for definiteness $ z $, which corresponds to wavelike and cosmological solutions in general relativity:
\be
ds^{2} = F (dt^{2} - dz^{2})- g_{ab} dx^{a}dx^{b},
\ee
where $ F = F (t, z) $, $ g_{ab} = g_{ab} (t, z) $, $ a, b = 1, 2 $ refer to the space variables $ x $ and $ y $. Let us pass to light-cone coordinates
\be
t = \zeta - \eta, \,\, z = \zeta + \eta
\ee
so that 
$$ F (dt^{2} - dz^{2}) = - 4 F(\zeta, \eta) \, d\zeta d\eta  $$ 
and write down equations (2) in these variables. We denote by $ g $ the $ 2 \times 2 $ matrix $ g_{ab} $ of space part of the metric tensor, and let 
$ \det g_{ab} = \alpha ^{2} $. Then equations (2) take the form:
\ba
\Big( \alpha  f_{R} M_{, \, \zeta} \, M^{-1}\Big)_{, \, \eta}  +   \Big( \alpha f_{R} M_{, \, \eta} \, M^{-1}\Big)_{, \,\zeta} & = & 0,\,\,\, M = \alpha^{-1}\, g \\
( \alpha  f_{R})_{, \,\, \zeta\eta} - \alpha F (f_{R} R - f) & = & 0 \\
\alpha  f_{R} F R + \alpha {f_{R}}_{, \,\, \zeta\eta} - 2 f_{R} {\alpha}_{, \,\, \zeta\eta} - \frac{1}{2} (\alpha_{, \,\, \eta} {f_{R}}_{, \,\, \zeta}  +  \alpha_{, \,\,\zeta} {f_{R}}_{, \,\,\eta}) & = & 0 
\ea
and two equations for metric coefficient $ F $ are:
\ba
(\ln F)_{, \,\,\zeta} = \dfrac{\ln ( \alpha  f_{R})_{, \,\, \zeta \zeta}}{\ln ( \alpha  f_{R})_{, \,\, \zeta}} - \dfrac{\textit{Tr} \, (g_{, \,\, \zeta} (g^{-1})_{, \,\, \zeta}) - 4 (\ln (f_{R})_{, \,\, \zeta})^{2}}{4 \ln ( \alpha  f_{R})_{, \,\, \zeta}}\\ \nonumber
(\ln F)_{, \,\,\eta} = \dfrac{\ln ( \alpha  f_{R})_{, \,\, \eta\eta}}{\ln ( \alpha  f_{R})_{, \,\, \eta}} - \dfrac{\textit{Tr} \, (g_{, \,\, \eta} (g^{-1})_{, \,\, \eta}) - 4 (\ln (f_{R})_{, \,\, \eta})^{2}}{4 \ln ( \alpha  f_{R})_{, \,\, \eta}}.
\ea
It is also useful to give the expression for $ (\ln F)_{, \,\zeta \,\eta} $:
\be
(\ln F)_{, \,\zeta \,\eta} = \dfrac{\alpha_{, \,\,\zeta} \alpha_{, \,\,\eta}}{\alpha^{2}} + \frac{1}{4} \textit{Tr} \, (g_{, \,\, \zeta} (g^{-1})_{, \,\, \eta}) - \frac{{f_{R}}_{, \,\, \zeta\eta}}{f_{R}} +  \dfrac{\alpha_{, \,\, \eta} {f_{R}}_{, \,\, \zeta}  +  \alpha_{, \,\,\zeta} {f_{R}}_{, \,\,\eta}}{2 \alpha  f_{R}}
\ee
although it is a consequence of the previous equations.

It is easy to see that, in contrast to the general relativity case \cite{BZ}, eqs. (8) are not independent. But the main difference between these equations and the equations considered in \cite{BZ} is as follows: despite the apparent similarity of matrix equation (5) and eq. (1.4) in \cite{BZ}, the L-A pair constructed in \cite{BZ} cannot be applied to (5) because of equation (6), or $ ( \alpha  f_{R})_{, \,\, \zeta\eta} \neq 0 $; the case of equality to zero gives $ f(R) = R $. However, it is not difficult to see that eq. (5) will not change if $ M \longrightarrow A M B $, where $ A $ and $ B $ are constant matrices with unit determinant (or $ A = B^{-1} $). This makes it possible to obtain, for example, the components $g_{xy} $ from the diagonal metric, but they will be expressed in terms of the diagonal components of the initial metric. In this paper we will restrict ourselves to considering the diagonal metric.

\section{Exact solution}

To find exact solutions of eqs. (5)-(8) we consider the diagonal metric in the form:
\ba
g_{ab} = \begin{pmatrix}
\alpha^{2s_{1}} & 0 \\
0 & \alpha^{2s_{2}}
\end{pmatrix}, \,\,\, s_{1} + s_{2} = 1.
\ea
Then eqs. (5) takes the form:
\ba
(2 s_{1} - 1) \Big ( 2 f_{R} \alpha_{, \,\, \zeta \eta} + \alpha_{, \,\, \eta} {f_{R}}_{, \,\, \zeta}  +  \alpha_{, \,\,\zeta} {f_{R}}_{, \,\,\eta} \Big ) & = & 0 \nonumber
\ea
and one can see that there are two possibilities: 
\ba
& Case \, \textbf{\textit{1}}.\,\,\,\,&  2 f_{R} \alpha_{, \,\, \zeta \eta} + \alpha_{, \,\, \eta} {f_{R}}_{, \,\, \zeta}  +  \alpha_{, \,\,\zeta} {f_{R}}_{, \,\,\eta} = 0, \,\, \, s_{1} \in [0, 1] \\
& Case \, \textbf{\textit{2}}.\,\,\,\,&  s_{1} = s_{2} = \frac{1}{2}
\ea 

Let us first consider the first case. 

\subsubsection{Case \textbf{\textit{1}}. }

Equation (7) becomes
\be
\alpha  f_{R} F R + (\alpha {f_{R}})_{, \,\, \zeta \eta} = 0
\ee
and together with eq. (6) this leads to the expression for $ f $:
\be
f = C_{R} \, \sqrt{\sigma_{R}\, R}, \,\,\, C_{R} = const, \,\,\, \sigma_{R} = \sign R.
\ee
Eqs. (8)-(9) take the following form:
\ba
(\ln F)_{, \,\, \zeta} & = & \dfrac{{\alpha}_{, \,\, \zeta \zeta} f_{R} + \alpha  {f_{R}}_{, \,\, \zeta \zeta }}{( \alpha f_{R})_{, \,\, \zeta}} \\ \nonumber
(\ln F)_{, \,\, \eta} & = & \dfrac{{\alpha}_{, \,\, \eta \eta} f_{R} + \alpha  {f_{R}}_{, \,\, \eta \eta }}{(\alpha f_{R})_{, \,\, \eta}} \\
(\ln F)_{, \,\, \zeta \eta} & = & - \dfrac{{f_{R}}_{, \,\, \zeta \eta}}{f_{R}} - \dfrac{(\alpha^{\gamma})_{,  \,\, \zeta \eta}}{\gamma \, \alpha^{\gamma}}, \,\,\, \gamma = 2 s_{1}^{2} - 2 s_{1} + 1 \neq 0.
\ea
Although the equations have been simplified, they are still difficult to solve without some additional conditions imposed on the functions. Let
\be
(\ln F)_{, \,\, \zeta \eta} = 0,
\ee
then $ F = F_{1}(\zeta)F_{2}(\eta) $. Further, suppose that $ \alpha =   {\alpha}_{1}(\zeta) {\alpha}_{2}(\eta) $ and $ f_{R} = {f_{R}}_{1}(\zeta){f_{R}}_{2}(\eta) $, besides from the latter equality it follows that $ |R| = |R_{1}(\zeta)| |R_{2}(\eta)| $. Then eqs. (15)-(16) give the relationship between $ f_{R} $ and $ \alpha $: 
\be
f_{R} = {f}_{0} {{\alpha}_{1}}^{\textit{k}}{{\alpha}_{2}}^{- \frac{\gamma}{{\textit{k}}}}, \,\,\,  {f}_{0} = const,
\ee
where $ \gamma = 1 $ only and ${\textit{k}} $  satisfies the equation $ {\textit{k}}^{2} + 2 {\textit{k}} - 1 = 0 $; we denote the roots of this equation as $ {\textit{k}}_{+} = \sqrt{2} - 1 $ and $ {\textit{k}}_{-} = -\sqrt{2} - 1 $. The condition $ \gamma = 1 $ means that $ s_{1} = 0, s_{2} = 1$ or vice versa, and we will consider the first case. It is also necessary to clarify that since $ f_{R} $ must be real, $ \alpha =   {\alpha}_{1}{\alpha}_{2} > 0 $.

So, we can now express all the variables in our model as functions of $ {\alpha}_{1} $ and of $ {\alpha}_{2} $, namely:
\ba
F & = & - \dfrac{2 \sigma_{R} {{f}_{0}^{2}}}{{C_{R}}^{2}} \, ({{\alpha}_{1}}^{2 \textit{k}})_{, \, \zeta}({{\alpha}_{2}}^{- \frac{2}{\textit{k}}})_{, \, \eta}\\
R & = & \dfrac{\sigma_{R}{C_{R}}^{2} }{4 {{f}_{0}^{2}} {{\alpha}_{1}}^{2 \textit{k}} {{\alpha}_{2}}^{- \frac{2}{\textit{k}}} }
\ea
provided that $ \sigma_{R} \sign(C_{R}{f}_{0} ) = 1 $. 

Consider now as an example of eq. (17) the simplest case of $ F $ in metric (5.17) in \cite{BZ} and compare the expressions for the metric interval. In the notation of \cite{BZ} and in coordinates $ \zeta, \eta $ this corresponds to $ m_{1} = b_{1} = 0$ and $ F = {a_{1}}^{2} \cosh(2\zeta) \cosh(2\eta) $. In our case, such a function $ F $ is possible for $ R > 0 $, and we obtain for $ \textit{k} = {\textit{k}}_{+} $ the interval:
\ba
ds^{2}  & = & - \dfrac{8 {{f}_{0}^{2}}}{{C_{R}}^{2}}  \cosh(2\zeta) \cosh(2\eta)\, d\zeta d\eta - dx^{2} - \frac{1}{4} \, \sinh(2\zeta)  \big( - \sinh(2\eta) \big) \Big(  \dfrac{\sinh(2\zeta) }{- \sinh(2\eta) }  \Big)^{\sqrt{2}}  dy^{2}.  \nonumber
\ea
In the initial coordinates $ t $ and $ z $ $ ds^{2} $ becomes
\ba
ds^{2}  & = &  \dfrac{2 {{f}_{0}^{2}}}{{C_{R}}^{2}} \, \cosh(t + z) \cosh(t - z)\, (dt^{2} - dz^{2}) - dx^{2} - \\  \nonumber
& - & \frac{1}{4} \, \sinh(t + z) \sinh(t - z)  \Big(  \dfrac{\sinh(t + z) }{ \sinh(t - z) }  \Big)^{\sqrt{2}}  dy^{2}; 
\ea
as follows from this equation, for $ t^{2} - z^{2} \geq 0 $ $ \alpha^{2} $ is real positive function. For this metric the scalar curvature is:
\be
R  =  \dfrac{{C_{R}}^{2}}{{f}_{0}^{2} \,  \sinh(t + z)  \, \sinh(t - z)   }
\ee
and $ R $ has a constant sign inside the entire light cone; on the generatrices of the light cone $ R \rightarrow + \infty $. 

If we take the new coordinates as
\be
\sinh(2\zeta)  =  \Big | \frac{C_{R}}{2f_{0}}\Big | (\tilde{t} + \tilde{z}), \,\,\,
\sinh(2\eta)  =  - \Big |\frac{C_{R}}{2f_{0}}\Big | (\tilde{t} - \tilde{z})
\ee
metric (21) takes the form
\ba
ds^{2}  =  d\tilde{t}^{2} - d\tilde{z}^{2} - dx^{2} - \frac{{C_{R}}^{2}}{8 {f}_{0}^{2}} \, (\tilde{t}^{2} - \tilde{z}^{2})\, \Big(  \dfrac{\tilde{t} + \tilde{z} }{ \tilde{t} - \tilde{z} }  \Big)^{\sqrt{2}}  dy^{2}.
\ea
The interval in \cite{BZ} has the form:
\be
ds^{2} =  {a_{1}}^{2} \cosh(t + z) \cosh(t - z) \, (dt^{2} - dz^{2}) - dx^{2} - {a_{1}}^{2} \cosh^{2}(t) \sinh^{2}(z) dy^{2},
\ee
and one can see that the intervals are very different. In new coordinates (23) ($|\frac{C_{R}}{2f_{0}}|$ must be replaced with $ \frac{1}{|a_{1}|}| $) we obtain from eq. (25) the flat interval:
\be
ds^{2} =  d \tilde{t}^{2} - d \tilde{z}^{2} - dx^{2} - \tilde{z}^{2} dy^{2}.
\ee
The question may arise when it is possible to take for the new coordinates $ {{\alpha}_{1}}^{2 \textit{k}}(\zeta) $ and $ {{\alpha}_{2}}^{- \frac{2}{\textit{k}}}(\eta) $ so that the new function $ F = 1 $. Such a change of coordinates has the form
\ba
{{\alpha}_{1}}^{2 \textit{k}}(\zeta) & = & \sqrt{\dfrac{{C_{R}}^{2}}{2 {{f}_{0}}^{2}}} \, (\tilde{\zeta}  + {{\alpha}_{1}}_{0} )  \\ \nonumber
{{\alpha}_{2}}^{- \frac{2}{\textit{k}}}(\eta) & = &  \sqrt{\dfrac{{C_{R}}^{2}}{2 {{f}_{0}}^{2}}} \, (\sigma_{R} \, \tilde{\eta}  + {{\alpha}_{2}}_{0}),  
\ea
$ {{\alpha}_{1}}_{0}$ and ${{\alpha}_{2}}_{0} $ are constant, and it can be seen that, for example, if both functions $ {{\alpha}_{1}}^{2 \textit{k}} $ and $ {{\alpha}_{2}}^{- \frac{2}{\textit{k}}} $ are positive then the new coordinates will not describe the past light cone. Therefore, the possibility of such a replacement must be analyzed in each specific case. Looking at eq. (23), we can assume that monotone odd functions may be suitable, but under an important condition: all variables of the model under consideration ($ \alpha $, $ g_{\mu\nu} $, $ f_{R} $) must be real.

\subsubsection{Case \textbf{\textit{2}}. }

First, similarly to Case $ \textbf{\textit{1}} $, we will obtain solutions for $ F $, $ \alpha $ and $ f_{R} $ in the form of a product of two functions, one of which depends only on $ \zeta $, and the other only on $ \eta $. Then eq. (9) gives
\be
\dfrac{\alpha_{, \,\,\zeta\eta}}{\alpha} - \frac{2 {f_{R}}_{, \,\, \zeta\eta}}{f_{R}} +  \dfrac{\alpha_{, \,\, \eta} {f_{R}}_{, \,\, \zeta}  +  \alpha_{, \,\,\zeta} {f_{R}}_{, \,\,\eta}}{\alpha  f_{R}} = 0,
\ee
and this has a solution in the form:
\ba
\alpha & = & {\alpha}_{1}(\zeta) \, {\alpha}_{2}(\eta) \\ \nonumber  
f_{R} & = & f_{0} \, {{\alpha}_{1}}^{p}  \,\,{{\alpha}_{2}}^{\frac{p+1}{2p-1}} , \,\,\, p \neq \frac{1}{2}, \,\,\, {f}_{0} = const. 
\ea
The cases $ p = 0 $ and $ p = -1 $ correspond to $ f \sim R $, that is  Einstein's gravity in vacuum, when $ R = 0 $. It can also be noted that eq. (29) turns into eq. (18) at $ p \neq {\textit{k}}_{\pm} $.

Further, substituting this into eqs. (6) and (7) we obtain
\ba
f & = & C_{R}\, |R|^{\frac{1-2p}{2p^{2} + 1}}, \,\,\, C_{R} = const \\
|R| & = & \sigma_{R} R = \Bigg(\dfrac{C_{R}\sigma_{R}(1-2p)}{f_{0} (2p^{2} + 1)}\Bigg)^{ \frac{2p^{2} + 1}{2p(p+1)}} \,\, \big({{\alpha}_{1}}^{\frac{2p^{2}+1}{2(p+1)}}\,{{\alpha}_{2}}^{\frac{2p^{2}+1}{2p(2p-1)}} \big)^{-1}.
\ea
It is easy to see that $ \frac{1-\sqrt{3}}{2} \leqslant \frac{1-2p}{2p^{2} + 1} \leqslant \frac{1+\sqrt{3}}{2} $. Thus, there are now more possibilities for the function $ f $ than in Case $ \textbf{\textit{1}} $. It can also to see that $ f \sim \sqrt{|R|} $ when $ p = - 1 \pm \frac{\sqrt{6}}{2} \neq  {\textit{k}}_{\pm}$.

It should be noted that the exponents here and below can take arbitrary values ​​(including fractional and irrational). Therefore, it is necessary to ensure that the values ​​raised to the power are positive for each particular $ p $ (except, possibly, integer $ p $). Finally, eqs. (15) take the following form:
\ba
(\ln F)_{, \,\, \zeta} & = & \dfrac{{\alpha}_{, \,\, \zeta \zeta} f_{R} + \alpha  {f_{R}}_{, \,\, \zeta \zeta }}{( \alpha f_{R})_{, \,\, \zeta}} - \frac{\alpha f_{R} }{2 ( \alpha f_{R})_{, \,\, \zeta}} \Big(\frac{{\alpha}_{, \,\, \zeta } }{\alpha}\Big)^{2}\\ \nonumber
(\ln F)_{, \,\, \eta} & = & \dfrac{{\alpha}_{, \,\, \eta \eta} f_{R} + \alpha  {f_{R}}_{, \,\, \eta \eta }}{(\alpha f_{R})_{, \,\, \eta}}- \frac{\alpha f_{R} }{2 ( \alpha f_{R})_{, \,\, \eta}} \Big(\frac{{\alpha}_{, \,\, \eta } }{\alpha}\Big)^{2} 
\ea
and we obtain the expression for $ F $:
\be
F = F_{0}  \big( {{\alpha}_{1}}^{\frac{2p^{2}+1}{2(p+1)}} \big)_{, \,\, \zeta} \,\, \big( {{\alpha}_{2}}^{\frac{2p^{2}+1}{2p(2p-1)}} \big)_{, \,\, \eta},
\ee
where 
\be
F_{0} = - \sigma_{R} \dfrac{6p(p+1)(1-2p)}{(2p^{2} + 1)^{2}} \Big(  \dfrac{\sigma_{R} f_{0} (2p^{2} + 1)}{C_{R}(1-2p)}\Big)^{ \frac{2p^{2} + 1}{2p(p+1)}}.
\ee
Again, as in Case $ \textbf{\textit{1}} $ the question arises under what conditions on the function $ {{\alpha}_{1}}^{\frac{2p^{2}+1}{2(p+1)}}(\zeta) $ and $ {{\alpha}_{2}}^{\frac{2p^{2}+1}{2p(2p-1)}}(\eta) $ it is possible to make $ F = 1 $. The choice of new coordinates is expressed similarly to eq. (27):
\ba
{{\alpha}_{1}}^{\frac{2p^{2}+1}{2(p+1)}}(\zeta) & = & \tilde{\zeta}  + {{\alpha}_{1}}_{0}   \\ \nonumber
{{\alpha}_{2}}^{\frac{2p^{2}+1}{2p(2p-1)}}(\eta) & = & - \frac{1}{F_{0}} \, ( \tilde{\eta}  + {{\alpha}_{2}}_{0}),  
\ea
$ {{\alpha}_{1}}_{0}$ and ${{\alpha}_{2}}_{0} $ are constant. But unlike Case $ \textbf{\textit{1}} $ both exponents of functions $ \alpha_{i} $ can be integer. So, their integer values $ \leq -5 $ and $ \geq 1 $. Let us consider both exponents equal to $ 1 $, or $ p = \frac{1 \pm \sqrt{3}}{2} $ and the trivial change of coordinates
\ba
{\alpha}_{1}(\zeta) & = & \tilde{\zeta}  =  \zeta = \frac{1}{2} (t + z)
\\ \nonumber
{\alpha}_{2}(\eta) & = & - \frac{\tilde{\eta}}{F_{0}}  =  \frac{\eta}{F_{0}}   = - \frac{1}{2 F_{0}} (t - z).
\ea
Then the interval becomes:
\be
ds^{2}  =  d{{t}}^{2} - d{{z}}^{2} + \frac{\sigma_{R}}{6 c_{0}} ({t}^{2} - {z}^{2} ) (dx^{2} + dy^{2})
\ee
where $ c_{0} = \Big(-\sigma_{R} \frac{f_{0} (1 \pm \sqrt{3})}{C_{R}}   \Big)^{\pm 2\sqrt{3}(2 \mp \sqrt{3})} $. It is immediately clear that $ \sigma_{R} = -1 $ and also $ \frac{f_{0} (1 \pm \sqrt{3})}{C_{R}} $ must be positive. The scalar curvature is
\be
R  =  - \dfrac{6}{{t}^{2} - {z}^{2}}, 
\ee
and as in eq. (22) $ R $ has a constant sign inside the entire light cone.   On the generatrices of the light cone $ R \rightarrow  - \infty $. 

Finally, let us obtain a class of solutions parameterized by two arbitrary functions for the case
\be
f  = \dfrac{C_{R}}{R}, \,\,\, C_{R} = - |C_{R}| = const;
\ee
obviously this case is not included in eq. (30). This is realized, in particular, under condition 
\be
\alpha (\zeta, \eta) = f_{R} (\zeta, \eta)
\ee
for which (after solving eqs. (6), (7) and (32)) we obtain solutions expressed in terms of two arbitrary functions $ {\varphi}_{1}(\zeta) $ and $ {\varphi}_{2}(\eta) $ each of which is not identically zero:
\ba
\alpha & = & f_{R} = \dfrac{9^{4}}{{C_{R}}^{2}} \,\, \Big(\int  {\varphi}_{1}(\zeta) d\zeta + \int  {\varphi}_{2}(\eta) d\eta   \Big)^{- 4} \\
F  & = & 4 \sigma_{R} \dfrac{ 9^{3} }{|C_{R}|^{\frac{3}{2}}} \,\,\, {\varphi}_{1}(\zeta) {\varphi}_{2}(\eta) \,\, \Big(\int  {\varphi}_{1}(\zeta) d\zeta + \int  {\varphi}_{2}(\eta) d\eta  \Big)^{-4} \\
R  & = & \frac{\sigma_{R} |C_{R}|^{\frac{3}{2}} }{9^{2}} \, \Big(\int  {\varphi}_{1}(\zeta) d\zeta + \int {\varphi}_{2}(\eta) d\eta   \Big)^{2}.
\ea
If it is possible to represent $ {\varphi}_{1}(\zeta) d\zeta \sim d\Phi_{1}(\zeta) $ and $ {\varphi}_{2}(\eta) d\eta \sim d\Phi_{2}(\eta) $ then the choice of coordinates 
\ba
\Phi_{1} & = & \frac{1}{2} (\tilde{t} + \tilde{z})
\\ \nonumber
\Phi_{2} & = & - \frac{9 \sigma_{R} }{8 |C_{R}|^{\frac{1}{2}}} \, (\tilde{t} - \tilde{z})
\ea
will bring the metric to a conformally flat form. As the simplest example one can take $ {\varphi}_{1}(\zeta) $ and $ {\varphi}_{2}(\eta) $ as constants and then we obtain a solution in terms of traveling wave variable.

\section{Conclusion}

In this paper we have found exact vacuum solutions for two-dimensional metric $f(R)$-gravity model in which all functions depend on time and one spacelike coordinate. The solutions obtained are expressed in terms of arbitrary functions. The possibility of choosing these functions as new variables is analyzed. 

The systems under consideration imply further development and generalization. For example, we can obtain exact solutions for two-dimensional stationary models and compare them with those obtained earlier by other authors.

\end{document}